\documentclass[5p]{elsarticle}
\usepackage{epsfig,psfrag,graphicx,bm,color,verbatim,amssymb,lineno}

\begin{document}

\begin{frontmatter}

\title{Dark Matter implications of the Fermi-LAT measurement of anisotropies in the diffuse gamma-ray background: status report}

\author[1,2]{Mattia Fornasa}
\author[3]{Jesus Zavala}
\author[4]{Miguel A. S\'{a}nchez-Conde}
\author[1]{Francisco Prada}
\author[5]{Mark Vogelsberger}

\address[1]{Instituto de Astrof\'isica de Andaluc\'ia - CSIC, Glorieta de la Astronom\'ia, E-18008, Granada, Spain}
\address[2]{MultiDark fellow}
\address[3]{Department of Physics and Astronomy, University of Waterloo, 200 University Avenue West, Waterloo, Canada}
\address[4]{KIPAC - SLAC National Accelerator Laboratory, 2575 Sand Hill Road, Menlo Park, CA 94025, USA}
\address[5]{Harvard-Smithsonian Center for Astrophysics, 60 Garden St., Cambridge, MA 02138, USA}

\begin{abstract}
For the first time, the Fermi-LAT measured the angular power spectrum (APS) 
of anisotropies in the diffuse gamma-ray background. The data is found to 
be broadly compatible with a model with contributions from the point sources 
in the 1-year catalog, the galactic diffuse background, and the extragalactic 
isotropic emission; however deviations are present at both large and small 
angular scales.
In this study, we complement the model with a contribution from Dark Matter 
(DM) whose distribution is modeled exploiting the results of the most recent 
$N$-body simulations, considering the contribution of extragalactic halos 
and subhalos (from Millenium-II) and of galactic substructures (from 
Aquarius). With the use of the Fermi Science Tools, these simulations serve 
as templates to produce mock gamma-ray count maps for DM gamma-ray emission, 
both in the case of an annihilating and a decaying DM candidate. The APS will 
then be computed and compared with the Fermi-LAT results to derive 
constraints on the DM particle physics properties. The possible systematic 
due to an imperfect model of the galactic foreground is also studied and 
taken into account properly.
The present paper reports on the status of the project.
\end{abstract}


\end{frontmatter}

\section{Introduction}
\label{sec:one}
The Isotropic Gamma-Ray Background (IGRB) can be defined as the radiation 
that remains after the resolved sources (both point-like and extended) 
and the galactic foreground (produced from the interaction of the cosmic rays
with the interstellar medium) are subtracted from the total gamma-ray 
emission. The most recent measurement of the IGRB energy spectrum was done
by the Fermi-LAT telescope and presented in 
Refs. \cite{Ackermann:2011,Abdo:2010nz}. 
Contrary to what was found by the EGRET telescope 
\cite{Sreekumar:1997un,Strong:2004de}, the IGRB appears to be perfectly 
compatible with a power-law with a slope of $-2.41$, at least up to 10 GeV. 

The contribution of unresolved sources, both extragalactic (e.g., 
blazars \cite{Stecker:1996ma,Pavlidou:2002va,Narumoto:2006qg,Ando:2006mt,
Ando:2006cr,SiegalGaskins:2009ux,Collaboration:2010gqa}, star-forming galaxies 
\cite{Pavlidou:2002va,Fields:2010bw} and radio galaxies \cite{Inoue:2011bm,
Inoue:2010iz}) and galactic sources (like milli-second pulsars 
\cite{FaucherGiguere:2009df}) have been considered. The contribution of each 
class is estimated from population studies of the detected objects (see 
Ref. \cite{Collaboration:2010gqa} for an example in the case of blazars) and 
it turns out that all the components considered up to now may not be able
to account for the whole IGRB \cite{Ajello:2011}. Thus there may be room for 
additional contributions like, e.g., that of Dark Matter (DM) 
\cite{Ullio:2002pj,Ando:2005xg,Ando:2006cr,Fornasa:2009qh,Abdo:2010dk}. 
Actually, the authors of Ref. \cite{Abdo:2010dk} have already used the 
Fermi-LAT measurement of the IGRB intensity to constrain the properties of 
the DM particle, finding that, in their most optimistic scenario, values 
larger than $10^{-26}$cm$^{3}$s$^{-1}$ for the annihilation cross section 
($\sigma_{\mbox{\tiny{ann}}}v$) can be excluded.

Complementary to the energy spectrum of the IGRB, we can also the recent
analysis of the angular power spectrum (APS) of anisotropies in the diffuse
gamma-ray background \cite{Vargas:2010en,SiegalGaskins:2010nh,
SiegalGaskins:2011} to study the nature of the IGRB. The analysis was 
conducted by the Fermi-LAT collaboration and resulted in the detection of 
some angular power above multipoles of $\ell>155$, with a significance 
larger than $3\sigma$, in the each of the energy bins between 1 GeV and 10 
GeV. More detailed information on this measurement will be provided in 
Sec. \ref{sec:two}. We simply note here that, even if such a detection seems 
to be compatible to the signal predicted for a population of unclustered, 
unresolved blazars, it can still be used to put some useful constraints on 
the nature of the DM particle. The goal of the present work is to derive 
those constraints using state-of-art numerical simuations.

The project is divided in two parts: in the first one we will make use of 
the most recents results from $N$-body simulations to derive all-sky maps 
of gamma-ray emission from DM annihilations and decay. We will consider 
the contributions of $i)$ extragalactic halos and subhalos (based on the 
Millennium II $N$-body simulation \cite{BoylanKolchin:2009nc}), $ii)$ the 
emission from the smooth halo of the Milky Way (modeled as in 
Ref. \cite{Prada:2004pi} and $iii)$ its subhalos (from the results of 
Aquarius \cite{Navarro:2008kc,Springel:2008cc}). Moreover, we will also
consider the emission from halos and subhalos below the mass resolution of 
the two simulations mentioned above, down to the minimal halo mass 
$M_{\mbox{\tiny{min}}}$. The all-sky maps will then be used to compute the APS 
of anisotropies in the gamma-ray emission from DM annihilation and decay. 
We plan to present the results of this first part in Ref. \cite{First_paper}, 
where we will also study the dependence of the APS from the assumptions made 
while building the maps of DM-induced gamma-ray emission. Refer also to
Secs. \ref{sec:three} and \ref{sec:four} for more information on this first 
part.

The second part of the project will deal with the comparison with Fermi-LAT 
APS data. The simulation maps produced in Ref. \cite{First_paper} will be 
used as templates for the DM contribution to the IGRB. We will use the most 
recent Fermi Instrument Response Functions (IRFs) to estimate how experimental 
issues may affect the APS. It will be particularly important to take 
correctly into account the experimental Point Spread Function (PSF) and the 
Fermi-LAT exposure. We will finally use the APS measurement to derive 
constraints on the nature of DM taking into account also the constraints 
from the new measurement of the IGRB energy spectrum (see also 
Sec. \ref{sec:five}).

We will conclude this small introduction reminding that this note should be 
considered just as a status report, since none of the two parts of the project 
is complete up to now. We will present some preliminary results in 
Secs. \ref{sec:three} and \ref{sec:four}, but we will mainly focus here on
the methodology, emphasizing that the final results will be extensively 
presented in the near future \cite{First_paper}.

\section{Fermi-LAT measurement of the anisotropies in the diffuse gamma-ray background}
\label{sec:two}
For the analysis in Ref. \cite{SiegalGaskins:2011} the first 22 months of 
Fermi-LAT data have been analyzed, dividing the energy range between 1 GeV 
and 50 GeV in 4 energy bins. The point sources in the first 1 year catalog 
\cite{Abdo:2010} have been masked, as well as the emission within a band of 
30 degrees above and below the galactic plane. This was done to cover the 
regions in the sky where the emission is dominated by resolved sources and by
the galactic foregound, and to restrict the analysis only to where the IGRB is 
a significant component. We note here that the analysis performed in
Ref. \cite{Abdo:2010nz} to measure the IGRB intensity spectrum is different 
since templates of emission are implemented to subtract known components
in the gamma-ray emission, instead than masking portions of the sky.
Thus, strictly speaking, the data used in Ref. \cite{SiegalGaskins:2011} and 
analyzed to derive the APS are not the same used for the IGRB energy spectrum  
in Ref. \cite{Abdo:2010nz}.
Moreover, even at high latitudes ($|l| \geq 30^\circ$) the galactic emission
is still important and non-negligeble. We therefore expect that some level of 
contamination in the APS may be present from this kind of background. These 
contaminations are supposed to be located primarly at low angular multipoles 
$\ell$ (large angular scales) and so only multipoles larger than 155 have been
considered in Ref. \cite{SiegalGaskins:2011}. On the other hand, multipoles 
larger than 504 are also discarded since, at these small angular scales, the 
signal is strongly damped by the experimental PSF.

Two definitions of APS are used: $i)$ the intensity APS as in 
Eqs. \ref{eqn:intensity_APS_1} and \ref{eqn:intensity_APS_2} below, for which 
the intensity gamma-ray maps are decomposed directly in spherical harmonics
\begin{equation}
a_{\ell,m} = \int I(\Psi) Y^\star_{\ell,m}(\Psi) d\Omega,
\label{eqn:intensity_APS_1}
\end{equation}
\begin{equation}
C_{\ell} = \sum_{-|\ell|}^{|\ell|} |a_{\ell,m}|^2,
\label{eqn:intensity_APS_2}
\end{equation}

and $ii)$ the fluctuaction APS, that can be derived by the intensity one, 
dividing for the average intensity squared.

As a consequence, the fluctuation APS will not depend on the energy of the
emission, if the distribution of the sources is energy-indenpendent.

The Fermi-LAT reported detection of angular power in all the 4 energy bins
considered, with a significance of larger than $3\sigma$ in the energy bins
from 1 GeV to 10 GeV. The data have been compared with the power spectrum
of a {\it source model} made of $i)$ the point-like sources in 
Ref. \cite{Abdo:2010}, $ii)$ a model for the galactic foreground and $iii)$ 
an isotropic component at the level of the IGRB in Ref. \cite{Abdo:2010nz}. 
In the same regions outside the mask defined above, the general features in
the APS of the model are similar to those in the data, but the model APS does
not accurately reproduce the data APS in all energy bins on small or large
angular scales. Furthermore, the model angular power at $155 \ge \ell \ge 504$ 
is consistently below that measured in the data.

This seems to point to the possibility of having a population of unresolved
sources contributing to the APS, in order to explain the data.
The fact that the intensity APS is compatible with being independent 
of multipole, together with the way the intensity and fluctuation APS
changes in the 4 energy bins suggests that one or more populations of 
unresolved, unclustered classes of sources may be responsible for the missing 
power.

If this hypothesis is true, then the data will provide us some useful 
constraints on the DM particle: the normalization and shape of the APS from 
DM annihilation has been studied extensively in the last years, focusing on 
the case of extragalactic halos and subhalos \cite{Ando:2005xg,Ando:2006cr,
Cuoco:2007id,Cuoco:2007sh,Zavala:2009zr}, on the case of a galactic component 
\cite{Ando:2009fp,SiegalGaskins:2008ge,SiegalGaskins:2009ux} or both 
\cite{Fornasa:2009qh,Cuoco:2010jb}. Comparing model predictions with the 
Fermi-LAT APS data allows us to put constraints on how important are DM-induced 
gamma-rays in the IGRB and, consequently, draw exclusion lines on the 
annihilation cross section, or other quantities more difficult to constrain 
like the minimal halo mass $M_{\mbox{\tiny{min}}}$.

\section{Extragalactic halos and subhalos}
\label{sec:three}
\begin{figure*}
\includegraphics[width=0.49\textwidth]{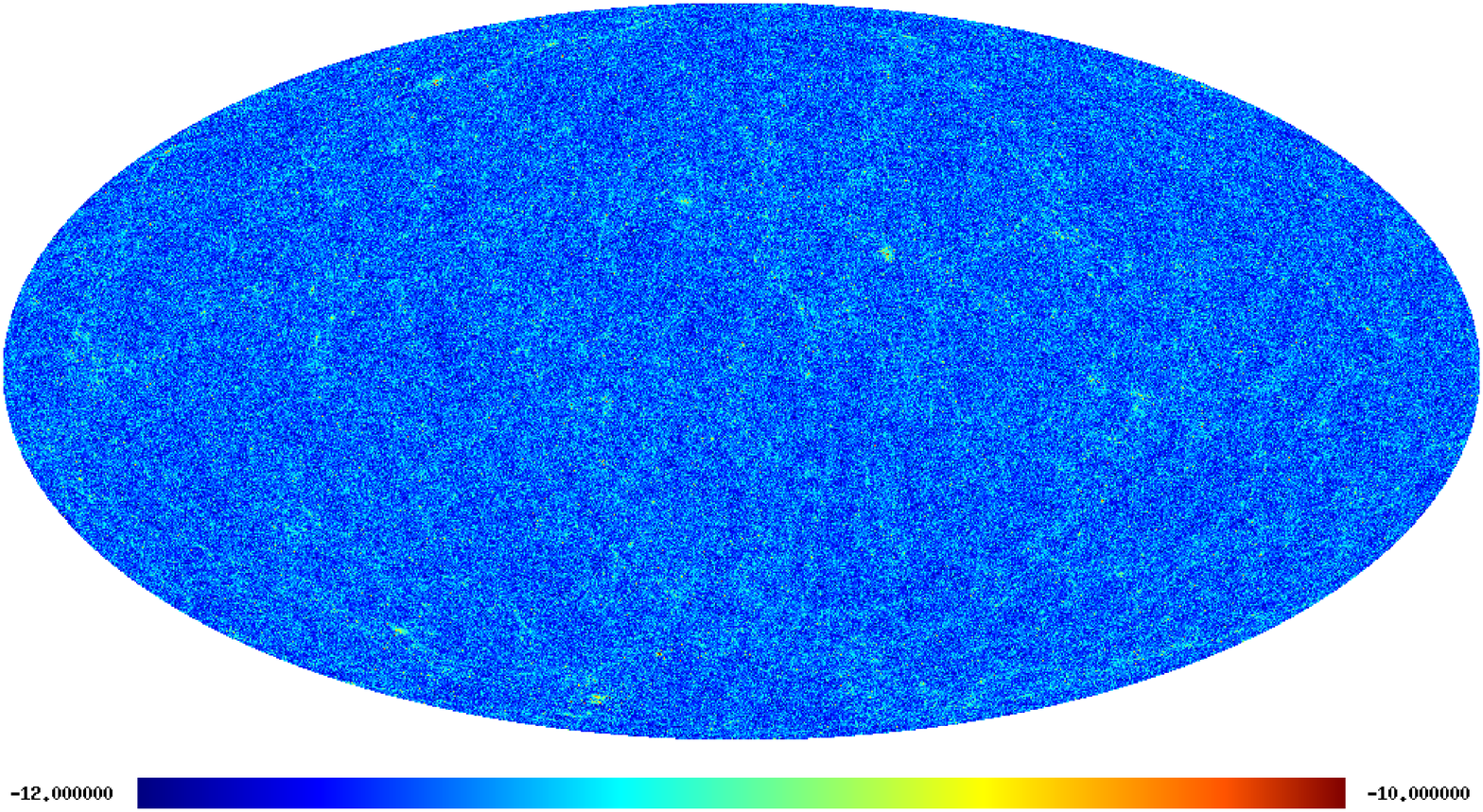}
\includegraphics[width=0.49\textwidth]{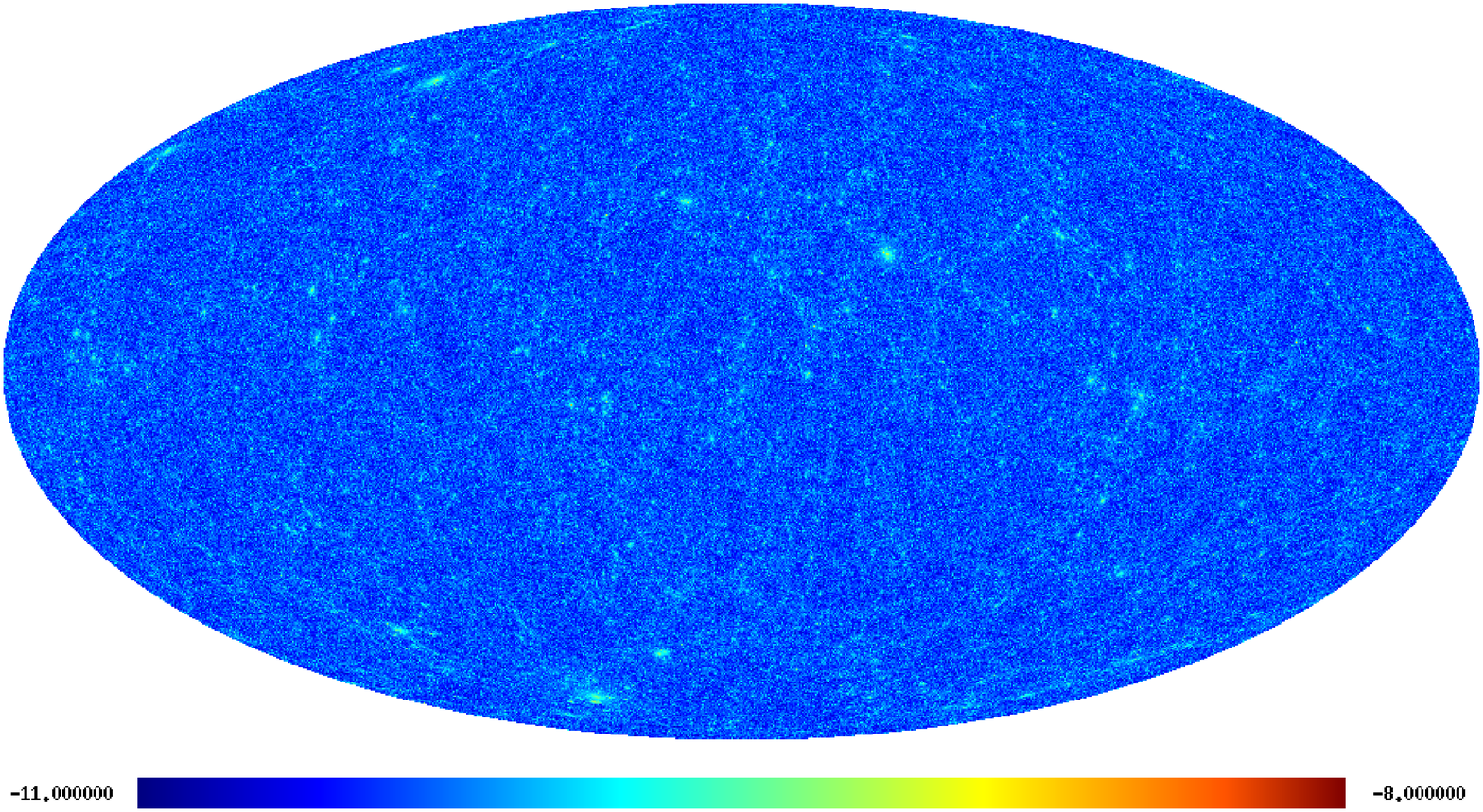}
\caption{\label{fig:Extragalactic_maps} All-sky maps of gamma-ray emission from DM annihilation (left panel) or decay (right panel) in halos and subhalos above the mass resolution of Millennium-II, up to $z=2.6$. Emission comes from hadronization of $b$ quarks and is computed at 10 GeV. For the case of annihilating DM, the mass is 200 GeV and the annihilation cross section $3 \times 10^{-26} \mbox{cm}^{3}\mbox{s}^{-1}$, while, for the decaying case, the mass is 2 TeV and the decay lifetime is $\tau=2 \times 10^{27} \mbox{s}$.}
\end{figure*}

This section and the next one will be devoted to describe the methodology
that will be used to compute the maps of DM-induced radiation.
To model the extragalactic emission, we use the catalogs of halos and 
subhalos of the Millennium-II $N$-body simulation: the simulation box is a 
cube with a size of 100 Mpc$/h$ and it contains DM halos and subhalos down
to a mass resolution of $M_{\mbox{\tiny{res}}}=6.89 \times 10^6 M_\odot/h$
\cite{BoylanKolchin:2009nc}. 
Halo catalogs are available at multiple snapshots with different redshifts
up to $z=127$.
We analyze them in a very similar way to what has been done in 
Ref. \cite{Zavala:2009zr}: we will only consider objects with more than 100
particles and derive the properties of each halo (DM profile, luminosity and
concentration) from the maximal circular velocity $V_{\mbox{\tiny{max}}}$ and
radius $r_{\mbox{\tiny{max}}}$ (corrected from spurious effects related to 
numerical resolution, as done in Ref. \cite{Zavala:2009zr}) and assuming a 
Navarro-Frenk-White (NFW) profile \cite{Navarro:1995iw}. 
Since we want to probe a volume which is much larger than the Millennium-II 
simulation box, we implement the same technique described in 
Ref. \cite{Zavala:2009zr} to construct sky-maps of the extragalactic signal, 
dividing the past light-cone in concentric shells (each of them at a 
particular redshift) and then filling them up with identical copies of the 
Millennium-II simulation boxes at that particular redshift (see Fig.9 in 
Ref. \cite{Zavala:2009zr}).
Each cube is randomly rotated and translated in order to avoid any
boundary effect. In this way one can compute the luminosity from a particular 
direction in the sky $\Psi$ simply by accounting for all the halos
encountered in the direction of $\Psi$.
This was done for all halos above Millennium-II mass resolution and you 
may see the results in Fig. \ref{fig:Extragalactic_maps} up to a $z=2.6$. 

We also want to consider the contribution of halos and subhalos below the
mass resolution of Millennium-II. We treated main halos and subhalos 
separetely and in two different ways.
We will start by describing how we account for main halos with a mass smaller
than $M_{\mbox{\tiny{res}}}$. Actually, the authors of Ref. \cite{Zavala:2009zr}
already considered this contribution: fitting the Millennium-II halos, they 
derived the cumulative luminosity function 
$F(M)=\sum L/\bar{M} \Delta\log M$, providing the total luminosity of main 
halos with a mass between $M$ and $M+dM$. Then, they assumed that $F(M)$ 
can be extrapolated below $M_{\mbox{\tiny{res}}}$ in order to compute the 
gamma-ray flux predicted from main halo between $M_{\mbox{\tiny{min}}}$ and 
$M_{\mbox{\tiny{res}}}$. This quantity was then used to boost up the luminosity 
of halos with a mass between $1.4 \times 10^8 M_\odot/h$ and 
$6.89 \times 10^8 M_\odot/h$. 
This was done under the assumption that the distribution of main halos below 
$M_{\mbox{\tiny{res}}}$ follows exactly that of those between 
$1.4 \times 10^8 M_\odot/h$ and $6.89 \times 10^8 M_\odot/h$, such an approach 
is by the fact that the two-point correlation function (which is an indicator
of clustering) is found to reach a constant value when approaching 
$M_{\mbox{\tiny{res}}}$ (see Fig. 10 of Ref. \cite{BoylanKolchin:2009nc}).

We want to test this assumption by implementing a different way of including 
the contribution of halos below $M_{\mbox{\tiny{res}}}$. We will then look for 
differences in the APS to check if the approach used to cover the range 
between $M_{\mbox{\tiny{min}}}$ and $M_{\mbox{\tiny{res}}}$ affects in any sense 
the shape of the APS. 
We therefore proceeded as follows:
\begin{itemize}
\item divide the range between $M_{\mbox{\tiny{min}}}$ and $M_{\mbox{\tiny{res}}}$
in mass decades $(M_i,M_{i+1})$. For each mass decade, we considered all the
halos in the Millennium-II simulation box with a mass between 
$1.4 \times 10^8 M_\odot/h$ and $6.89 \times 10^8 M_\odot/h$ and assign to each
of them a new mass between $M_{i}$ and $M_{i+1}$, assuming masses follow a 
probability distribution equal to the halo mass function $dn/dM$.
\item we also assign a luminosity to each of these halos.
From these two quantities, we can completely derive the halo profile.
In this way, we end up by having a box of simulated objects with 
masses between $M_i$ and $M_{i+1}$ distributed as those between in the range 
$1.4 \times 10^8 - 6.89 \times 10^8 M_\odot/h$.
\item however, in the mass decades we are considering, the halo mass function
will predict more halos than we have actually have in the simulated box. So,
we stack together different copies of the box obtained in the previous
point until we reach the correct number of halos. The different copies
are randomly rotated one with respect to the other.
\item the box that comes out from the stacking is used to fill up the
region up to a maximal distance of $R_{\mbox{\tiny{max}}}$, defined as the
distance above which all halos in a given mass decade become point-like. Maps
are produced from this distribution of objects till $R_{\mbox{\tiny{max}}}$.
\item for the regime beyond $R_{\mbox{\tiny{max}}}$, we do not consider any
stacking, we simply take the boxes coming from point 2 of this list and 
create the sky-map from those. Multiple copies of the map are then stacked 
together till the desidered flux is reached. Since we are now in the regime 
where halos are point-like, the procedure is similar to what has been done
below $R_{\mbox{\tiny{max}}}$ but without having to consider the 3D distribution 
of halos.
\end{itemize}

This procedure, in a sense, moves in the opposite direction to what has
been done in Ref. \cite{Zavala:2009zr} since we start by assuming that halos
below $M_{\mbox{\tiny{res}}}$ have the same distribution of those above, but
then we pass through a lot of independent rotations so that, at the end,
the final maps are expected to be more isotropic than those in 
Ref. \cite{Zavala:2009zr}. Once we compute the APS from the two approaches 
we will be able to see if differences show up in the shape of the APS.

In Fig. \ref{fig:IGRB} (left panel) the black crosses indicate the IGRB
energy spectrum taken from Ref. \cite{Abdo:2010nz}. The red points
show the amount of flux produced in halos and subhalos above
$M_{\mbox{\tiny{res}}}$ until $z=2.6$ (those depicted in 
Fig. \ref{fig:Extragalactic_maps}) in the case of annihilating DM (filled 
points) and decaying DM (empty points). If we also consider the emission 
below $M_{\mbox{\tiny{res}}}$ (implemented as described in the list above) 
the emission increases to be that one of the green points. For annihilating 
DM the increase is approximately of an order of magnitude, while for decaying 
DM it is completely negligible (red and green empty points practically 
overlap).
The total flux is still a factor of 100 (50) for annihilation (decay) below
the IGRB, at least at 10 GeV. 

In order to complete the description of the extragalactic component, the only
remaining ingredient is the emission from subhalos below $M_{\mbox{\tiny{res}}}$.
Those will be described following the prescription derived in 
Ref. \cite{Kamionkowski:2010mi} and generalized in 
Ref. \cite{SanchezConde:2011ap}. In the first reference the subhalos of the
Via Lactea II $N$-body simulation are analyzed to compute the probability
distribution $P(\rho,r)$ of having a DM density between $\rho$ and 
$\rho+d\rho$ at a distance $r$ from the center of the MW halo. $P(\rho,r)$
has two distinct components: $i)$ a parabolic regime due to the smooth halo 
and $ii)$ a power-law that extends to larger densities due to substructures 
(see Fig. 1 of Ref. \cite{Kamionkowski:2010mi}).
The authors, then used $P(\rho,r)$ to compute the boost factor due to
substructures, while Ref. \cite{SanchezConde:2011ap} extended the 
prescription to halos larger and smaller than the MW halos, having in mind
the case of galaxy clusters and dwarf Spheroidal galaxies respectively.
We will use it to boost up the emission of the halos below 
$M_{\mbox{\tiny{res}}}$, accounting in this way for the emission of unresolved
subhalos.
 
\section{The Milky Way halo and its subhalos}
\label{sec:four}
We continue now with the implementation of the gamma-ray emission related to 
our own galaxy: we use the distance of 700 kpc (approximately 3 times the 
virial radius of the Milky Way (MW) halo) as the limit between the 
extragalactic and the galactic regime.

The smooth halo of the MW is modeled with a NFW profile with parameters taken
from Ref. \cite{Prada:2004pi}, a model that is consistent with the available 
observation data on the MW. The total flux of the MW smooth halo is plotted 
in the left panel of Fig. \ref{fig:IGRB}. It represents the largest among 
the different DM components plotted there. However we should note that at 
the moment of computing the APS, we will mask the galactic plane, as it 
has been done by the Fermi-LAT collaboration in Ref. \cite{SiegalGaskins:2011}.
This will have the effect of drastically decreasing the emission associated
with the smooth halo.

To include the galactic subhalos we will use the Aquarius $N$-body simulation 
\cite{Springel:2008cc}. The same procedure described in the previous section 
will be used here for the Aquarius subhalos, without the need to considering 
replicas of the simulation box since we are interested only in the region 
within 700 kpc. The emission associated with galactic subhalos is indicated as 
yellow points in Fig. \ref{fig:IGRB}. We have not yet implemented the 
contribution of subhalos below the mass resolution of Aquarius 
$M_{\mbox{\tiny{res}}}^\prime=1.712 \times 10^3 M_\odot$. 
This will substantially increase the emission associated with galactic 
subhalos, which is only subdominant if we only consider objects above 
$M_{\mbox{\tiny{res}}}^\prime$.

\section{Deriving constraints}
\label{sec:five}
\begin{figure*}
\includegraphics[width=0.32\textwidth]{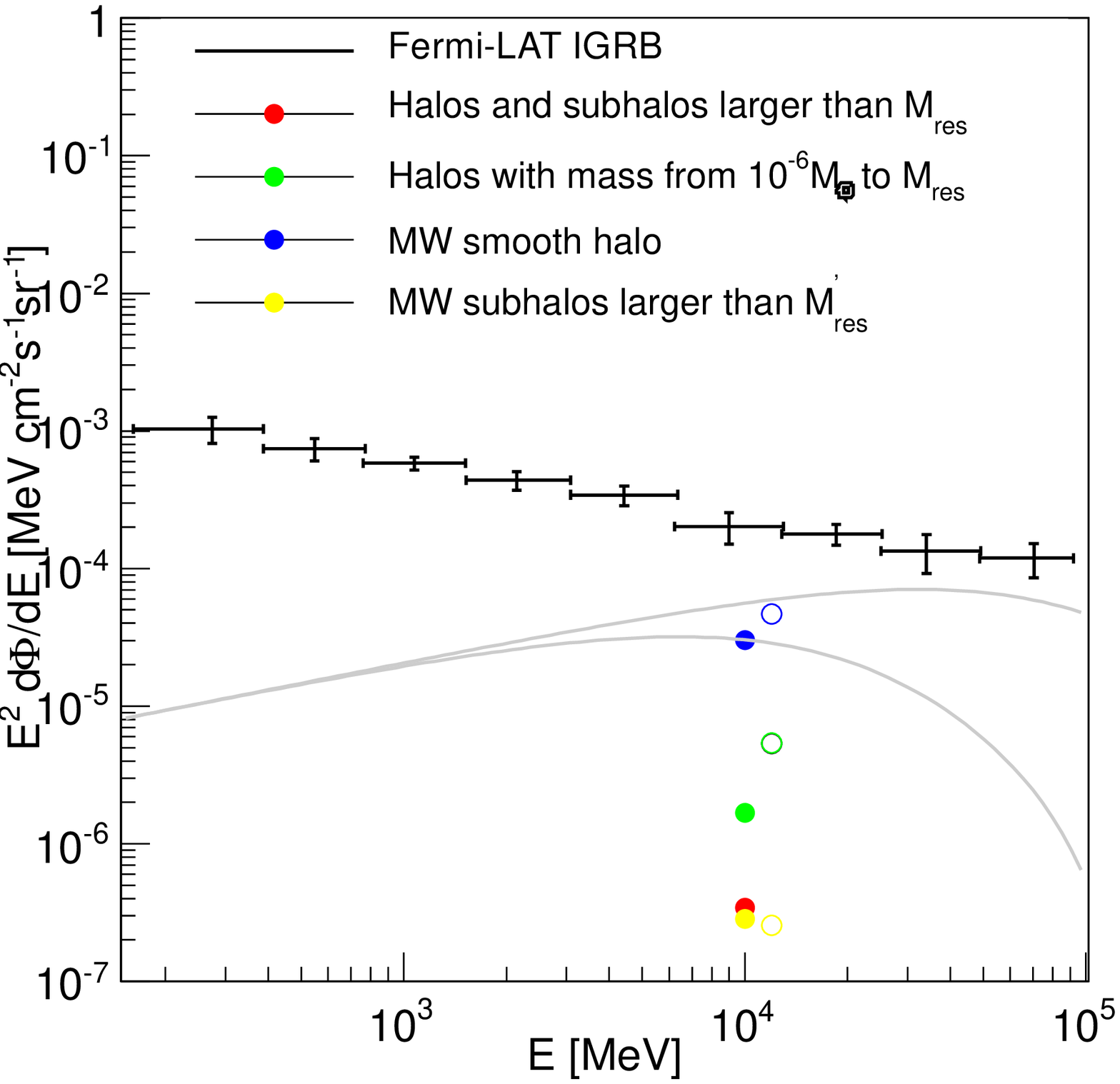}
\includegraphics[width=0.32\textwidth]{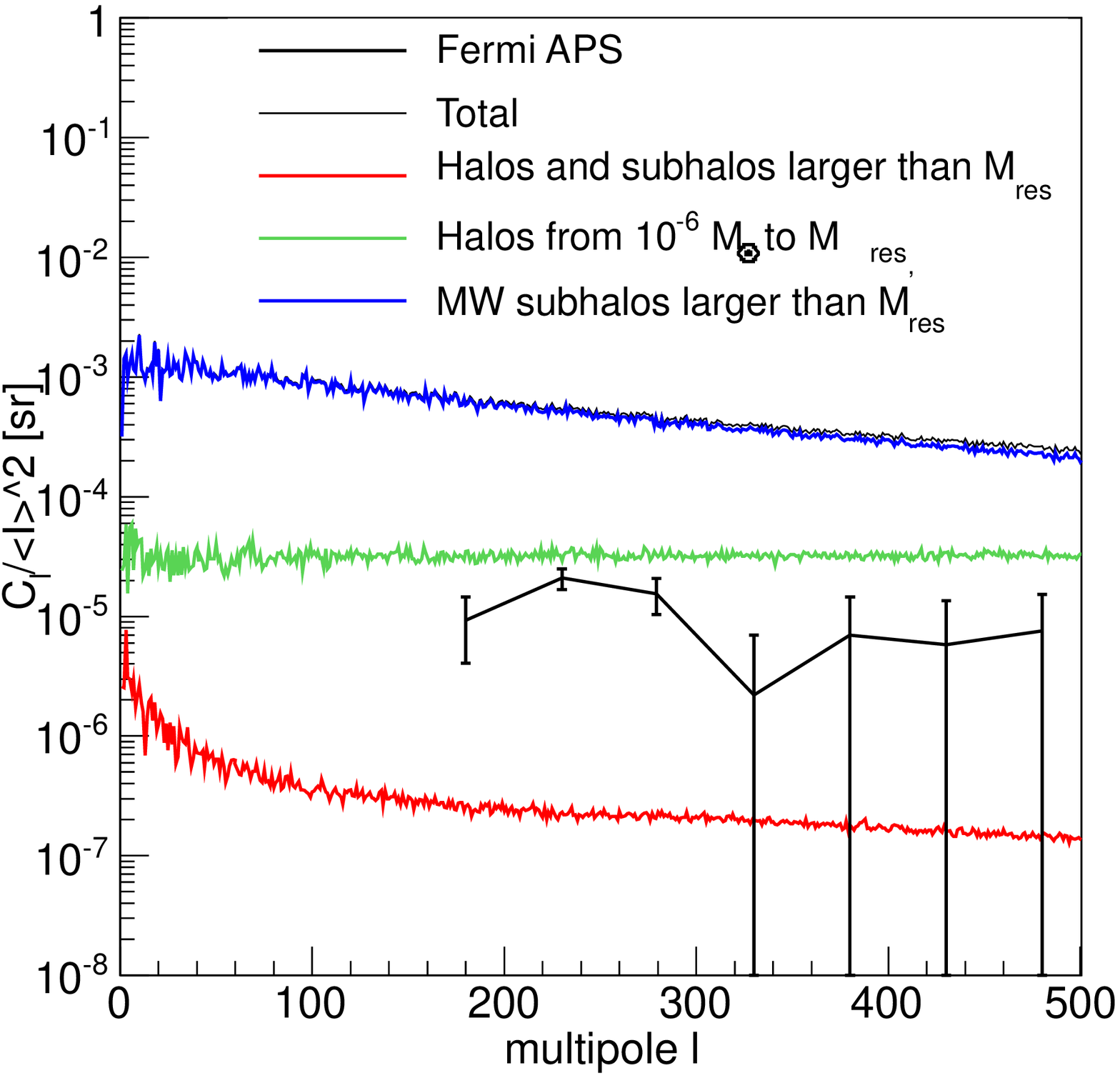}
\includegraphics[width=0.32\textwidth]{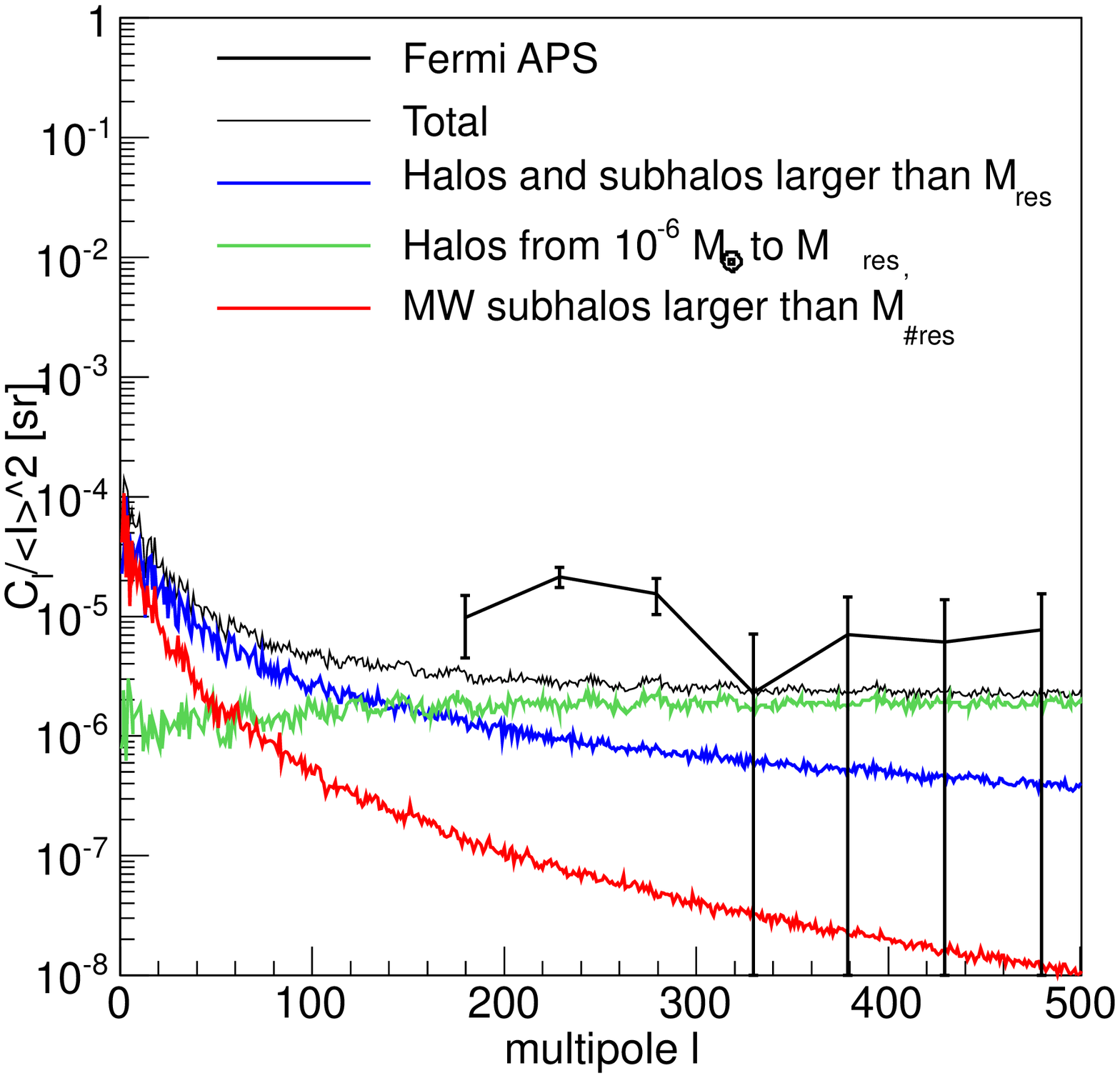}
\caption{\label{fig:IGRB} Left panel: Black crosses indicate the energy spectrum of the IGRB taken from Ref. \cite{Abdo:2010nz}. Filled (empty) points refer to the case of annihilating (decaying) DM. Red points indicate the flux coming from DM halos and subhalos above $M_{\mbox{\tiny{res}}}$, while green points only consider the extrapolation down to $M_{\mbox{\tiny{min}}}$ for main halos. Blue points show the emission from the smooth DM halo of the Milky Way, while the yellow ones accounts for the emission from all the subhalos in the Aquarius simulation. The grey lines refer to the annihilation and decay energy spectrum (from the hadronization of $b$ quarks) normalized to blue points. Central and right panel: APS of anisotropies in the gamma-ray emission from DM annihilation (central panel) and decay (right panel). The total fluctuation APS is plotted in black, while the colored lines correspond to the different components, multiplied by the square of their average flux with respect to the total average flux. The red line indicates the contribution of extragalactic halos and subhalos above the mass resolution of Millennium II. The green line refers to main halos below $M_{\mbox{\tiny{res}}}$ and down to $M_{\mbox{\tiny{min}}}$, while the blue one accounts for the subhalos in the Aquarius simulation.}
\end{figure*}

The maps that will be obtained from the procedure sketched in the previous
sections will then be used to compute the APS and to compare the results with
the Fermi-LAT data. Some results can be seen in the central and right panels
of Fig. \ref{fig:IGRB} where the total fluctuation APS is plotted (black 
line). Also the contribution of the different components is present, 
multiplied by the square of the average flux of each component with respect 
to the total average emission, so that the sum of the colored lines gives 
the total (black line) also visually.
Let us stress that these are very preliminary results: not all the components 
have been included (the contribution of unresolved extragalactic and 
galactic subhalos is missing) and we are not considering any mask.

We can, however, already see some interesting trends:
\begin{itemize}
\item the galactic component dominates the case of annihilating DM. This 
may change after we will include the contribution of galactic subhalos 
less massive than $M_{\mbox{\tiny{res}}}^\prime$.
\item the extragalactic component below $M_{\mbox{\tiny{res}}}$ (both for 
annihilating and decaying DM) does not indicate any intrinsic clustering, 
as expected.
\item the contributions of Aquarius subhalos (blue lines) and of 
Millennium-II halos and subhalos (red lines) decrease with multipole, a
consequence of the fact that the APS is sensitive to the inner structures
of the halos. 
\end{itemize}

In Fig. \ref{fig:IGRB} we have also plotted the Fermi-LAT measurement of 
the fluctuation APS in the energy bin between 1 GeV and 2 GeV. Some other
ingredients are still needed in order to be able to perform a proper 
comparison between the DM APS and the data. This will be done in the future,
taking into account experimental features like the effect of the PSF,
or a possible residual contamination from the Galactic foreground. We plan 
to include the IRF of the telescope using the Fermi Science Tools and 
treating the DM component in the same way as one of the three different 
components of the source model described in Sec. \ref{sec:two}.

\section{Conclusions}
\label{sec:six}
Here we briefly summarized a project which is still in progress. The goal is
to compute realistic and complete all-sky maps of gamma-ray emission from DM
annihilation and decay, taking into account the contribution of both
extragalactic and galactic halos and subhalos. For the most massive objects
we will refer directly to the results of two of the most recent $N$-body
simulations (Millennium-II and Aquarius), but we will also account for
smaller halos, down to their minimal mass $M_{\mbox{\tiny{min}}}$. Once
obtained, these maps will be used to compute the APS of anisotropies. Our
goal is to compare it with the recent measurement of the APS by the Fermi-LAT
collaboration in order to be able to put constraints on the nature of
the DM particle.
As the project is still on-going, we only presented here the methodology 
and some preliminary results. We refer interested readers to keep an eye on 
the arXiv webpage for the publication of the final results 
\cite{First_paper}.



\section*{Acknowledgements}
The simulations used in this paper were carried out in the Cosmology Machine
supercomputer at the Institute for Computational Cosmology, Durham. The 
Cosmology Machine is part of the DiRAC facility jointly funded by STFC, the
Large Facilities Capital Fund of BIS and Durham University. We would also 
like to thank the Virgo Consortium for giving us access to the data of the
Millnnium-II and Aquarius simulations. Finally we thank the support of the 
Consolider-Ingenio 2010 Programme under grant Multi-Dark CSD2009-00064.


\end{document}